\documentclass{article}
\usepackage{spconf, amsmath, graphicx, hyperref, xcolor, subcaption, braket, amsthm, amssymb, nicefrac}
\usepackage{algorithm} 
\usepackage{algpseudocode}

\def\x{{\mathbf x}}

\makeatletter
\def\thmheadbrackets#1#2#3{%
  \thmname{#1}\thmnumber{\@ifnotempty{#1}{ }\@upn{#2}}%
  \thmnote{ {\the\thm@notefont(#3)}}}
\makeatother

\newtheoremstyle{brakets}
  {}
  {}
  {\itshape}
  {}
  {\bfseries}
  {.}
  { }
  {\thmheadbrackets{#1}{#2}{#3}}

\theoremstyle{brakets}
\newtheorem{theorem}{Theorem}

\newtheorem{proposition}{Proposition}
\newtheorem{assumption}{Assumption}
\newtheorem{definition}{Definition}

\title{Stochastic Shadow Descent: Training Parametrized Quantum Circuits with Shadows of Gradients}
\name{Sayantan Pramanik$^{1,2}$ \qquad M Girish Chandra$^{1}$}
\address{$^{1}$TCS Research, TATA Consultancy Services, $^{2}$Indian Institute of Science \\ \{sayantan.pramanik, m.gchandra\}@tcs.com}
\begin{document}
\newcommand{\Or}{\mathcal{O}}
\newcommand{\loss}{\mathcal{L}}
\newcommand{\U}{\mathbf{U}}
\newcommand{\V}{\mathbf{V}}
\newcommand{\W}{\mathcal{W}}
\newcommand{\Ham}{\mathbf{H}}
\newcommand{\Op}{\mathbf{O}}
\newcommand{\ran}{\mathbf{v}}
\newcommand{\rant}{\ran^{(t)}}
\newcommand{\dd}{D_\ran(\z)}
\newcommand{\ddt}{D^{(t)}_\ran(\z)}
\newcommand{\ddcap}{\hat{D}_\ran(\z)}
\newcommand{\ddtcap}{\hat{D}^{(t)}_\ran(\z)}
\newcommand{\z}{\boldsymbol{\theta}}
\newcommand{\zt}{\boldsymbol{\theta}^{(t)}}
\newcommand{\ztp}{\boldsymbol{\theta}^{(t+1)}}
\newcommand{\nf}{\hat{f}\big(\z\big)}
\newcommand{\gp}{g_{\ran}^{\mu}}
\newcommand{\f}{f\big(\z\big)}
\newcommand{\ft}{f\big(\zt\big)}
\newcommand{\ftp}{f\big(\ztp\big)}
\newcommand{\F}{F\left(\z\right)}
\newcommand{\Ft}{F\left(\zt\right)}
\newcommand{\Ftp}{F\left(\ztp\right)}
\newcommand{\fs}{f^*}
\newcommand{\df}{\f-\fs}
\newcommand{\dft}{\ft-\fs}
\newcommand{\dftp}{\ftp-\fs}
\newcommand{\edf}{\mathbb{E}\left[\df\right]}
\newcommand{\edft}{\mathbb{E}\left[\dft\right]}
\newcommand{\edftp}{\mathbb{E}\left[\dftp\right]}
\newcommand{\eft}{\mathbb{E}\left[\ft \right]}
\newcommand{\eftp}{\mathbb{E}\left[\ftp \right]}
\newcommand{\eF}{\mathbb{E}\left[\F \right]}
\newcommand{\eFt}{\mathbb{E}\left[\Ft \right]}
\newcommand{\eFtp}{\mathbb{E}\left[\Ftp \right]}
\newcommand{\cedf}{\mathbb{E}\left[\df \vert \mathcal{F}_t \right]}
\newcommand{\cedft}{\mathbb{E}\left[\dft \big\vert \mathcal{F}_t \right]}
\newcommand{\cedftp}{\mathbb{E}\left[\dftp \big\vert \mathcal{F}_t \right]}
\newcommand{\ceF}{\mathbb{E}\left[\F \vert \mathcal{F}_t \right]}
\newcommand{\ceFt}{\mathbb{E}\left[\Ft \vert \mathcal{F}_t \right]}
\newcommand{\ceFtp}{\mathbb{E}\left[\Ftp \vert \mathcal{F}_t \right]}
\newcommand{\gt}{\boldsymbol{g}^{(t)}}
\newcommand{\gf}{\nabla\f}
\newcommand{\gft}{\nabla\ft}
\newcommand{\gftp}{\nabla\ftp}
\newcommand{\lb}{\left(}
\newcommand{\rb}{\right)}
\newcommand{\at}{\alpha_t}
\newcommand{\egt}{\mathbb{E} \left[ \lvert\lvert \nabla \ft \rvert\rvert^2 \right]}
\newcommand{\br}{\mathbb{R}}
\newcommand{\brd}{\mathbb{R}^d}
\newcommand{\WRP}{\par\qquad\(\hookrightarrow\)\enspace}
\maketitle
\begin{abstract}
In this paper, we focus on the task of optimizing the parameters in Parametrized Quantum Circuits (PQCs). While popular algorithms, such as Simultaneous Perturbation Stochastic Approximation (SPSA), limit the number of circuit-execution to two per iteration, irrespective of the number of parameters in the circuit, they have their own challenges. These methods use central-differences to calculate biased estimates of directional derivatives. We show, both theoretically and numerically, that this may lead to instabilities in \emph{training} the PQCs. To remedy this, we propose Stochastic Shadow Descent (\texttt{SSD}), which uses random-projections (or \emph{shadows}) of the gradient to update the parameters iteratively. We eliminate the bias in directional derivatives by employing the Parameter-Shift Rule, along with techniques from Quantum Signal Processing, to construct a quantum circuit that parsimoniously computes \emph{unbiased estimates} of directional derivatives. Finally, we prove the convergence of the \texttt{SSD} algorithm, provide worst-case bounds on the number of iterations, and numerically demonstrate its efficacy.
\end{abstract}
\section{Introduction}
\label{sec:intro}
While Variational Quantum Algorithms (VQAs) \cite{vqe, vqa} are poised to have a major impact in Chemistry \cite{vqe, Kandala2017-yt}, Machine Learning \cite{Cerezo_qml, schuld_benchmarking}, and Optimization \cite{qaoa, qopt}, optimizing the parameters in the Parametrized Quantum Circuits (PQCs) remains a significant challenge \cite{amira}. This is partly due to the inaccessibility of intermediate quantum states \cite{amira}, and partly due to the presence of barren plateaus \cite{bp_review}. While the issue of unavailability of gradients was resolved by the Parameter-Shift Rules (PSR) \cite{psr, mitarai}, obviating the need for finite-differences, they pose new scaling challenges. The number of circuit-executions required to estimate gradients scales linearly with the number of parameters \cite{amira}. The use of Simultaneous Perturbation Stochastic Approximation (SPSA) \cite{spsa1, spsa2}, and its variants, has been suggested as a potential fix \cite{ibm_spsa, qnspsa}. SPSA uses two circuit-executions and the central-difference method to essentially find biased estimators of the directional derivative at every iteration of the algorithm. Akin to PSR, we construct a circuit which can be executed twice to get unbiased estimators of the directional derivative, eliminating the need for central-differences, while retaining the favorable scaling of SPSA-like zeroth-order methods.

The paper is organized as follows: we formally state the problem-setting and assumptions in Section \ref{sec:setting}, summarize our contributions in Section \ref{sec:contributions}, and detail our motivations in Section \ref{sec:motivation}. In Section \ref{sec:dd}, we construct quantum circuits which can be used to compute directional derivatives directly, and introduce the Stochastic Shadow Descent algorithm in Section \ref{sec:ssd}. Finally, we test the effectiveness of the algorithm in Section \ref{sec:exp}, and conclude in Section \ref{sec:conclusions}. The proofs of propositions and theorems have been excluded in adherence to the page-limit.

\subsection{Problem-Setting and Assumptions}\label{sec:setting}
In the context of VQAs, the objective function is given by:
\begin{equation}\label{eq:obj}
    \min_{\z \in \brd} \left( \f:= \braket{\psi(\z)|\Ham|\psi(\z)} \right);
\end{equation}
where $\ket{\psi(\z)} = \prod_{j=1}^d \V_j \U_j(\z_j)\ket{\iota}$ is the variational state parametrized by $\z \in \brd$, with $\ket{\iota}$ being the initial state; and $\Ham$ is a Hermitian observable. As a result, the objective values are real, for all $\z$. Further, similar to \cite{zoe}, the $\V_j$s are fixed, non-parametrized unitaries, and $\U_j = e^{-i\z_j\boldsymbol{\sigma}_j}$ for all $j \in [d]$. Also, $\boldsymbol{\sigma}_j$s are the Pauli matrices, and all $\z_j$s are uncorrelated with each other. For the rest of this paper, we make the following assumptions:
\begin{assumption}\label{as:smooth}
    The objective function $f$ bounded below by $\fs$ on $\brd$. Further, $f$ is $L$-smooth w.r.t. $\z$, i.e., $\forall \; \x, \mathbf{y} \in \brd$, $||\nabla f(\mathbf{y})- \nabla f(\x)||_2 \leq L ||\mathbf{y}-\x||_2$.
\end{assumption}
\begin{assumption}\label{as:noise}
    We have access only to noisy estimates $\nf$ of $\f$, such that $\mathbb{E}[\nf] = \f$, and $\mathbb{E}[(\nf-\f)^2] \leq \sigma^2$, for all $\z \in \brd$. 
\end{assumption}
\noindent The objective of the classical optimization algorithms is to reach an $\varepsilon$-stationary solution (in expectation) to Problem \ref{eq:obj}, which is formally defined as follows:
\begin{definition}
    A point is called an $\varepsilon$-stationary solution in the mean-squared sense if $\mathbb{E}[||\nabla \f||^2] \leq \varepsilon^2$.
\end{definition}

\section{Main Contributions}\label{sec:contributions}
First, we propose a method that can be used to obtain \emph{unbiased estimates} of directional derivatives of the parameters in PQCs. This method, outlined in Section \ref{sec:dd}, involves the execution of two circuits - termed Inner Product Circuits (\texttt{IPC}s) - with a marginal increase in depth and ancillary qubits (which is logarithmic in the number of parameters). An example of such a circuit has been shown in Fig. \ref{fig:dd2}. We further extend this method by constructing a second circuit (exemplified in Fig. \ref{fig:dd1}) which needs to be evaluated only once to estimate the directional derivative. This new circuit requires only one extra ancillary qubit over the \texttt{IPC}. 
    
Next, we introduce an iterative algorithm called Stochastic Shadow Descent (\texttt{SSD}) (Algorithm \ref{alg:ssd} in Section \ref{sec:ssd}) which uses directional derivatives (computed using the aforementioned circuits) to optimize the parameters in the PQCs. We also prove its convergence, along with finding upper bounds on the number of iterations required to obtain an $\varepsilon$-stationary solution to the objective function in \eqref{eq:obj} (Theorem \ref{th:ssd}).

\section{Motivation}\label{sec:motivation}
Perhaps the most ubiquitous algorithm used to optimize the parameters in PQCs is Stochastic Gradient Descent (SGD) \cite{sgd_handbook, sgd}. However, it can be shown that SGD requires $\Or(\nicefrac{Ld}{\varepsilon^4})$ iterations and execution of $\Or(\nicefrac{Ld^2}{\varepsilon^4})$ circuits to reach an $\varepsilon$-stationary solution to Problem \ref{eq:obj}. This quadratic dependence on $d$ hinders the scalability and inclusion of more parameters in the circuit \cite{amira}. It stems from the use of the PSR \cite{psr, gpsr} to estimate the gradients at each iteration. For the setting in Section \ref{sec:setting}, the partial derivative corresponding to the $i^{\text{th}}$ coordinate is given by $\nabla_i \f = \nicefrac{1}{2}\left( f\left(\z + \frac{\pi}{2}\mathbf{e}_i \right) -  f\left(\z - \frac{\pi}{2}\mathbf{e}_i \right) \right)$, where $\mathbf{e}_i$ is the unit vector along the $i^{\text{th}}$ coordinate. It is easy to infer that constructing the full gradient vector requires $2d$ circuit-executions.

\begin{figure}
    \centering
    \includegraphics[width=\linewidth]{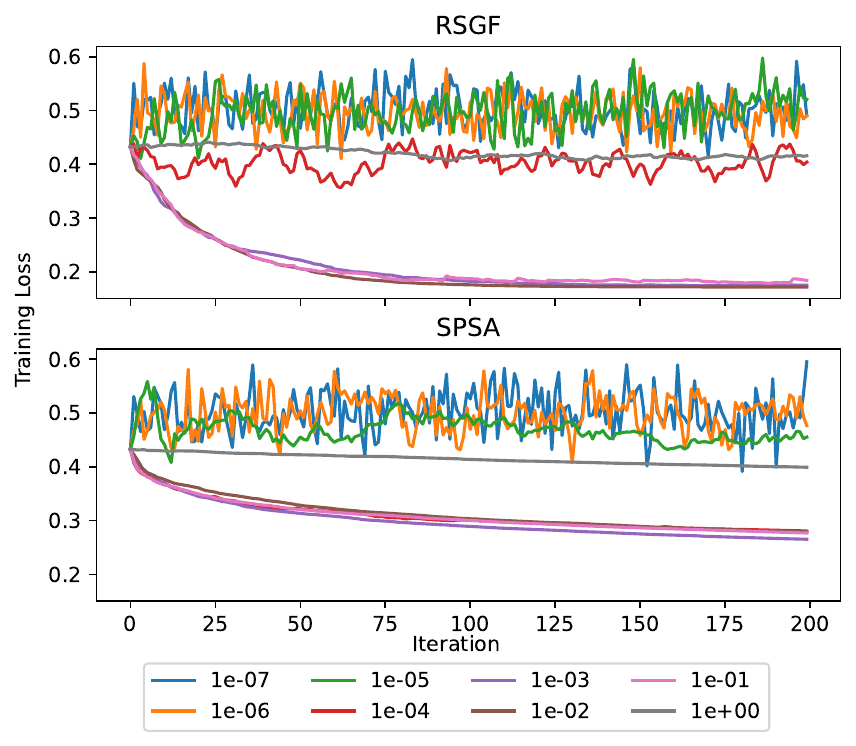}
    \caption{Plots depicting the training loss incurred while training a simple quantum model on the Iris dataset with RSGF and SPSA methods, respectively, for various values of $\mu$.}
    \label{fig:perturbations}
\end{figure}

The use of SPSA, which consumes only two circuit-executions per iteration, has been proposed as an alternative to SGD. For ease of analysis, we ignore the effect of noise temporarily and focus on the closely-related Randomized Stochastic Gradient Free (RSGF) \cite{Nesterov, Ghadimi_Lan} method. It uses a random vector $\rant \sim \mathcal{N}(\mathbf{0}, \mathbf{I})$ and a perturbative factor $\mu >0$ to compute the forward difference $g(\zt) = \nicefrac{1}{\mu}(f(\zt+\mu\rant)-f(\zt))$ at each iteration $t$.
If the step-size is $\alpha_t >0$, then the parameters are updated as: $\ztp = \zt -\alpha_t g(\zt)\rant$. It is clear from Eq. 3.16 of \cite{Ghadimi_Lan} that after running $T$ iterations of RSGF, $\min_{t\in [T]}\mathbb{E}[||\nabla \ft||^2]$ is upper bounded by $\Or(\mu^2d^2L)$ for an optimal selection of the step-size as per Eq. 3.24. Further, according to Eq. 3.25, an upper bound on $\mu$ requires knowledge of $L$ and $\fs$. This suggests that the algorithm may achieve its best performance for vanishingly small values of $\mu$. However, in the presence of noise, small values of $\mu$ may lead to instabilities during the optimization process. To demonstrate this, we trained a simple quantum classifier (detailed in \cite{itrust}) to distinguish between the first two classes of the Iris dataset with the SPSA and RSGF algorithms, and different scales of $\mu$. The training plots are shown in Fig. \ref{fig:perturbations}.

Finally, we define the (unnormalized) directional derivative of $f$ at $\z$ along $\ran$ as $\dd := \braket{\nabla \f, \ran}$, and propose that if $\mu \rightarrow 0$, $g(\z) = \dd$, and $\mathbb{E}[g(\z)\ran] = \nabla \f$:
\begin{proposition}\label{prop:dd}
    If Assumption \ref{as:smooth} is true, then $\forall\; \z, \ran \in \brd$: 
    
    $\underset{\mu \rightarrow 0}{\lim} \, \frac{1}{\mu}\left(f(\z+\mu\ran)-f(\z)\right) = \braket{\nabla \f, \ran}$.
    
    \noindent Further, if $\ran \sim \mathcal{N}(\mathbf{0}, \mathbf{I})$, then: 
    
    $\underset{\ran}{\mathbb{E}} \left[ \underset{\mu \rightarrow 0}{\lim} \,\frac{1}{\mu}\left(f(\z+\mu\ran)-f(\z)\right)\ran \right] = \nabla \f$.
\end{proposition}
\noindent Estimating the $\dd$ with PSR would also require the execution of $2d$ circuits. This gives us the motivation to directly compute the directional derivatives with a constant number of circuit-executions, and to eliminate the hyperparameter $\mu$ from SPSA and RSGF.

\section{Estimating Directional Derivatives with Quantum Circuits}\label{sec:dd}
In this section, we present a quantum circuit which can be used to directly compute directional derivatives without having to find the full gradients. This method, thus, requires the execution of only 2, instead of $2d$ circuits. The circuit is constructed using a combination of the PSR \cite{psr}, and a modification of the Linear Combination of Unitaries approach \cite{lcu}, which is a technique from Quantum Signal Processing.  First, we first express the directional derivative $\dd$, at $\z$ along the vector $\ran$, in a more convenient form. Recalling that $\dd = \sum_i \nabla_i \f \ran_i$, we replace the partial derivatives with the corresponding expression from the PSR \cite{psr}, such that $\dd$ can be written as the sum of two terms, with positive and negative shifts by $\nicefrac{\pi}{2}$, respectively:
\begin{equation}\label{eq:dd_ipc}
    \begin{aligned}
    \dd &= \frac{1}{2} \left(\sum_{i=1}^d f\left(\z + \frac{\pi}{2}\mathbf{e}_i \right)\ran_i - \sum_{i=1}^d f\left(\z - \frac{\pi}{2}\mathbf{e}_i \right)\ran_i \right) \\
        &= \frac{d}{2} \left(D^{+}_\ran (\z)-D^{-}_\ran (\z)\right).
\end{aligned}
\end{equation}
Here, the quantities $D^{\pm}_\ran (\z)$ are obtained from the quantum circuits, and are defined as follows:
\begin{equation}
    D^{\pm}_\ran (\z) := \frac{1}{d}\sum_{i=1}^d f\left(\z\pm \frac{\pi}{2}\boldsymbol{e}_i \right) \ran_i.
\end{equation}

\begin{figure}
    \centering
    \includegraphics[width=\linewidth]{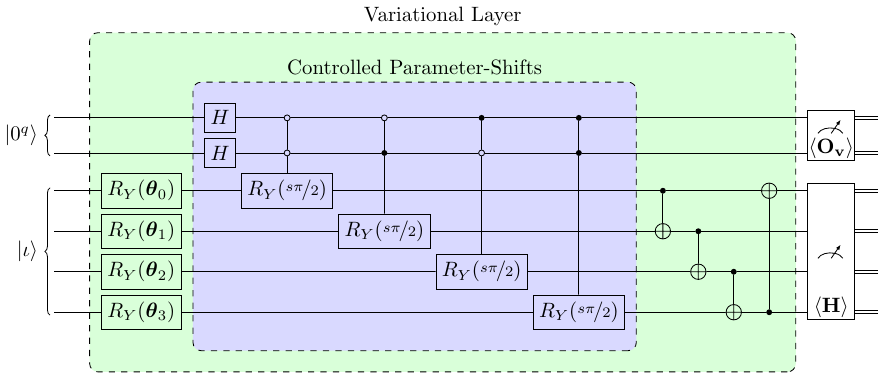}
    \caption{The \texttt{IPC} (constructed using Algorithm \ref{alg:ipc}) corresponding to a single variational layer of the \texttt{BasicEntanglerLayers} ansatz with 4 qubits and 4 parameters. The circuit returns $D^s_\ran (\z)$ as an output.}
    \label{fig:dd2}
\end{figure}
\begin{figure*}
    \centering
    \includegraphics[width=0.8\linewidth]{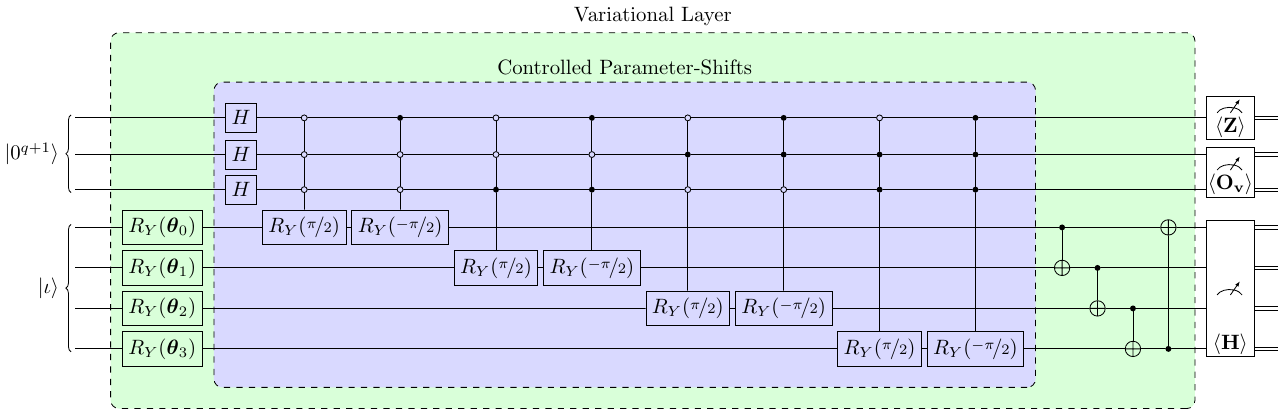}
    \caption{An enhanced version of the \texttt{IPC} depicted in Fig. \ref{fig:dd2}, which can be executed just once to compute $\dd$. The output of this circuit is $\nicefrac{\left(D^{+}_\ran (\z)-D^{-}_\ran (\z)\right)}{2}$. Here, $\mathbf{Z}$ is the Pauli-Z matrix.}
    \label{fig:dd1}
\end{figure*}

The procedure to construct the proposed quantum circuits - dubbed Inner Product circuits (\texttt{IPC}s) - which return the values of $D^{\pm}_\ran (\z)$, has been described in Algorithm \ref{alg:ipc}. The algorithm takes $\z$, $\ran$, and $s \in \{\pm1\}$ as inputs, and returns the corresponding $D^{\pm}_\ran (\z)$, which can then be used to calculate $\dd$ as per Eq. \eqref{eq:dd_ipc}. This requires only two circuits to be executed. It is noteworthy that although the requisite number of circuit-evaluations has been reduced by a factor of $d$, the \texttt{IPC}s described in Algorithm \ref{alg:ipc} require a larger number of qubits and have an increased depth compared to the original PQC. Evidently, the circuits need $q = \lceil\log_2{d}\rceil$ ancillary qubits. While the increase in qubits can be quantified easily, the increase in depth that comes from the controlled-unitaries is more nuanced, which depends on the topology of qubit-connectivity in the quantum processor and its native set of gates.
\begin{algorithm}
	\caption{Inner Product Circuit \texttt{(IPC)}} 
    \label{alg:ipc}
	\begin{algorithmic}[1]
            \State \textbf{Inputs:} Parameters $\z \in \brd$, vector $\ran \in \brd$, $s \in \{\pm 1\}$.
		\State Initialize the main register to the state $\ket{\iota}$
                \State Prepare the state $\sum_{i=0}^{d-1}\nicefrac{\ket{i}}{\sqrt{d}}$ in the ancillary register
			\For {$i \in [d]$}
                    \State Apply the variational gate $\U_i(\z_i)$
                    \State \parbox[t]{\dimexpr\linewidth-\algorithmicindent}{Apply the gate $\U_i(\nicefrac{s\pi}{2})$ controlled on the state $\ket{i}$ of the auxiliary qubits}
                    \State Apply the fixed gate $\V_i$
                \EndFor
            \Comment{Let the resulting state be denoted by $\ket{\phi(\z)}$}
            \State Construct the observable $\Op_\ran = \sum_{i=1}^d \ran_i \ket{i}\bra{i}$
			\State $D^s_\ran (\z) \leftarrow \braket{\phi(\z)|\Op_\ran \otimes \Ham|\phi(\z)}$ \Comment{Here, the observables $\Op_\ran$ and $\Ham$ are corresponding to the ancillary and main registers, respectively.}
        \State \textbf{Output:} $D^s_\ran (\z)$
	\end{algorithmic} 
\end{algorithm}
Fig. \ref{fig:dd2} illustrates the \texttt{IPC} for a single variational layer of the \texttt{BasicEntanglerLayers} ansatz from Pennylane as an example. The number of circuit-executions can be further reduced from two to one through the inclusion of another ancillary qubit and extra controlled gates, as shown in Fig. \ref{fig:dd1}. The output of this circuit is $\nicefrac{1}{2}\left(D^{+}_\ran (\z)-D^{-}_\ran (\z)\right)$, which can then be used to find $\dd$.

\section{Stochastic Shadow Descent}\label{sec:ssd}
Having discussed how two (or one) executions of a circuit can be used to directly compute the directional derivatives, we now outline the Stochastic Shadow Descent algorithm, which uses the said directional derivatives to iteratively update the parameters in PQCs. The algorithm is similar to RSGF \cite{Ghadimi_Lan} and SPSA \cite{spsa1} in that it replaces the biased gradient estimates along a random direction, with unbiased directional derivatives (or \emph{shadows} of the gradient). At each iteration $t$ of the \texttt{SSD} algorithm, a random direction $\rant$ is sampled from the standard normal distribution. Next, the directional derivative $\ddt$ at $\zt$ along $\rant$ is estimated using the circuits described in Section \ref{sec:dd}. Finally, the parameters are updated with a step-size $\alpha_t > 0$ as 
$\ztp = \zt - \alpha_t \ddtcap \rant$. The effect of noise (including shot-noise) is considered through the following assumption:
\begin{assumption}\label{as:dd}
    If the estimated directional derivative $\dd$ of $f$ at $\z$ along $\ran$ is denoted by $\ddcap$, then $\forall \; \z, \ran \in \brd$, $\mathbb{E}[\ddcap] = \dd$, and $\mathbb{E}[ (\ddcap-\dd)^2]\leq \eta^2$.
\end{assumption}
\noindent The \texttt{SSD} method has been formally stated in Algorithm \ref{alg:ssd}.

\begin{algorithm}
	\caption{Stochastic Shadow Descent \texttt{(SSD)} Algorithm} 
    \label{alg:ssd}
	\begin{algorithmic}[1]
            \State \textbf{Inputs:} Initial parameters $\z^{(0)}\in \mathbb{R}^d$, Sequence of step-sizes $\{\alpha_t\}$, Number of iterations $T>0$.
		\For {$t \in [T]$}
                \State $\rant \sim \mathcal{N}(\mathbf{0}, \mathbf{I})$
                \State $\hat{D}_+ \leftarrow \texttt{IPC}(\zt, \rant, s=1)$ \Comment{Using Algorithm \ref{alg:ipc}}
                \State $\hat{D}_- \leftarrow \texttt{IPC}(\zt, \rant, s=-1)$ \Comment{Using Algorithm \ref{alg:ipc}}
                \State $\ddtcap \leftarrow \frac{d}{2}\left(\hat{D}_+-\hat{D}_-\right)$
			\State $\ztp \leftarrow \zt -\alpha_t \ddtcap \rant$
		\EndFor
        \State \textbf{Output:} Sequence $\{\zt\}$
	\end{algorithmic} 
\end{algorithm}

\subsection{Convergence of \texttt{SSD}}
Now, we show that the \texttt{SSD} algorithm converges to an $\varepsilon$-stationary solution with the execution of $\Or(\nicefrac{Ld}{\varepsilon^4})$ circuits:
\begin{theorem}\label{th:ssd}
    Under Assumptions \ref{as:smooth} and \ref{as:dd}, if Algorithm \ref{alg:ssd} is executed for $T\geq 1$ iterations with a fixed step-size $\alpha \leq \nicefrac{1}{L(d+2)}$, then:
    
    \noindent $\min_{t \in [T]} \mathbb{E}[||\nabla \ft||^2] \leq \nicefrac{2}{\alpha T}(f(\z^{(0)})-\fs) + L \alpha d \eta^2.$
    
    \noindent Consequently, it can be guaranteed that for every $\varepsilon>0$, $\min_{t \in [T]} \mathbb{E}[||\nabla \ft||^2] \leq \varepsilon^2$ if $\alpha \leq \min \left\{  \nicefrac{1}{L(d+2)}, \nicefrac{\varepsilon^2}{2Ld\eta^2} \right\}$ and $T\geq \nicefrac{4(f(\z^{(0)})-\fs)}{\varepsilon^2} \max \left\{ L(d+2), \nicefrac{2Ld\eta^2}{\varepsilon^2}\right\}$.
\end{theorem}
\noindent Since each iteration of \texttt{SSD} requires (at most) two circuits to be evaluated, the total number of circuit-executions is $2T$.

\section{Experiments and Results}\label{sec:exp}
We tested the efficacy of the \texttt{SSD} algorithm by training a quantum classifier to differentiate between the digits $3$ and $5$ of the downscaled-MNIST dataset \cite{schuld_benchmarking}. The dataset $\{(\x_k, y_k)\}$ consisted of 100 randomly sampled training examples, with $\x_k \in \br^{10}$, and $y_k \in \{\pm 1\}$. The input features were standardized and encoded into the states of $10$ qubits with a combination of Hadamard and $R_Z$ gates, followed by $4$ layers of the \texttt{StronglyEntanglingLayers} ansatz from Pennylane, resulting in a circuit with $120$ parameters. Finally, the first qubit was measured in the Pauli-Z basis. If the measured value for the $k^{\text{th}}$ example is denoted by $f(\x_k, \z) \in [-1, 1]$, then the loss function can be written as $\nicefrac{1}{2}(1-y_k f(\x_k, \z))$. 

\begin{figure}
    \centering
    \includegraphics[width=\linewidth]{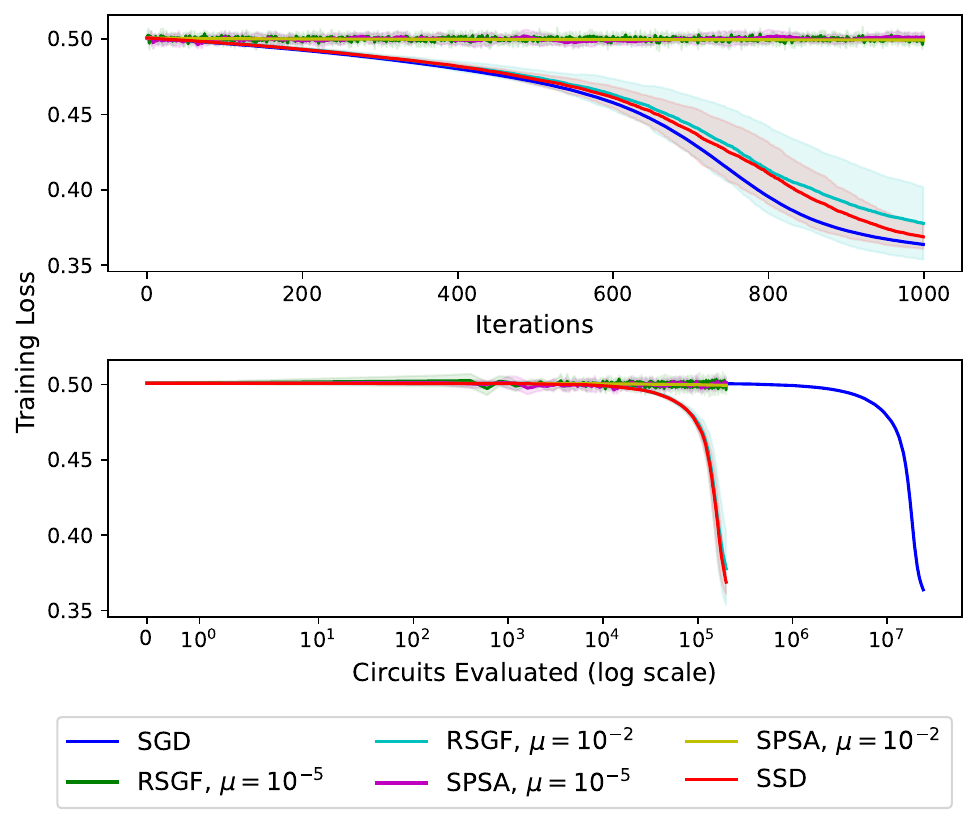}
    \caption{Plot of training loss against iterations, and the number of circuit-executions for SGD, RSGF, SPSA, and \texttt{SSD}.}
    \label{fig:results}
\end{figure}

The performance of \texttt{SSD} was compared against that of SGD, RSGF, and SPSA, all of which shared a common learning rate of $0.1$. RSGF and SPSA were executed with $\mu = 10^{-5}$ (calculated using Eq. 3.25 of \cite{Ghadimi_Lan}), and $\mu = 10^{-2}$ (found by performing a grid-search over a range of values). The other hyperparameters of SPSA were set as per the suggestions in \cite{spsa2}. The results of the experiment have been depicted in Fig. \ref{fig:results}, where the upper plot shows the training loss against the number of iterations, and the lower plot portrays the same against the total number of circuits evaluated. We note that the performance of \texttt{SSD} is on par with that of SGD (but with  a $\approx 100\times$ reduction in circuit-executions), which is matched by RSGF only after proper tuning of $\mu$.

\section{Conclusions}\label{sec:conclusions}
In this paper, we presented a method to directly estimate directional derivatives using quantum circuits, and demonstrated an effective way to leverage them to drastically reduce circuit-executions during PQC-training. More importantly, we showed that the use of standard, out-of-the-box classical optimization algorithms is not the best choice for VQAs. Instead, the algorithms can be judiciously adapted to address the unique constraints of quantum computation. We hope that this work will further stimulate an interest in the research and development of \emph{quantum-aware} optimization algorithms for learning-systems.

\bibliographystyle{IEEEbib}
\bibliography{refs}

\end{document}